\documentclass[sigconf]{acmart}
\usepackage{hyperref}
\usepackage{booktabs}
\usepackage{multirow}
\usepackage{makecell}

\usepackage{amsmath}
\usepackage{multirow}
\usepackage{booktabs}

\usepackage[linesnumbered,ruled,vlined]{algorithm2e}
\usepackage{subfigure}
\usepackage{soul}
\usepackage{xcolor}
\definecolor{newgreen}{RGB}{34,139,34}

\usepackage{caption}
\usepackage{subcaption}
\AtBeginDocument{%
  \providecommand\BibTeX{{%
    \normalfont B\kern-0.5em{\scshape i\kern-0.25em b}\kern-0.8em\TeX}}}

\acmSubmissionID{0}

\begin{document}

\title{Investigating Effective Speaker Property Privacy Protection in Federated Learning for Speech Emotion Recognition}


\author{Chao Tan}
\affiliation{%
  \institution{MonotaRO Co., Ltd.}
  \city{Osaka}
  \country{Japan}}
\email{chaotan2003@gmail.com}
\orcid{0009-0009-1978-7812}

\author{Sheng Li}
\affiliation{%
  \institution{NICT}
  \city{Kyoto}
  \country{Japan}
}\email{sheng.li@nict.go.jp}
\orcid{0000-0001-7636-3797}

\author{Yang Cao}
\affiliation{%
 \institution{Tokyo Institute of Technology}
 \city{Tokyo}
 \country{Japan}}
 \email{cao@c.titech.ac.jp}
 \orcid{0000-0002-6424-8633}

\author{Zhao Ren}
\affiliation{%
  \institution{University of Bremen}
    \city{Bremen}
  \country{Germany}
}\email{zren@uni-bremen.de}
\orcid{0000-0003-0707-5016}

\author{Tanja Schultz}
\affiliation{%
  \institution{University of Bremen}
    \city{Bremen}
  \country{Germany}
}\email{tanja.schultz@uni-bremen.de}
\orcid{0000-0002-9809-7028}

\renewcommand{\shortauthors}{Tan and Li et al.}

\begin{abstract}
Federated Learning (FL) is a privacy-preserving approach that allows servers to aggregate distributed models transmitted from local clients rather than training on user data. More recently, FL has been applied to Speech Emotion Recognition (SER) for secure human-computer interaction applications. Recent research has found that FL is still vulnerable to inference attacks. To this end, this paper focuses on investigating the security of FL for SER concerning property inference attacks. We propose a novel method to protect the property information in speech data by decomposing various properties in the sound and adding perturbations to these properties. Our experiments show that the proposed method offers better privacy-utility trade-offs than existing methods. The trade-offs enable more effective attack prevention while maintaining similar FL utility levels. This work can guide future work on privacy protection methods in speech processing.
\end{abstract}

\begin{CCSXML}
<ccs2012>
<concept>
<concept_id>10003120.10003121</concept_id>
<concept_desc>Human-centered computing~Human computer interaction (HCI)</concept_desc>
<concept_significance>500</concept_significance>
</concept>
<concept>
<concept_id>10002978</concept_id>
<concept_desc>Security and privacy</concept_desc>
<concept_significance>500</concept_significance>
</concept>
</ccs2012>
\end{CCSXML}

\ccsdesc[500]{Human-centered computing~Human computer interaction (HCI)}
\ccsdesc[500]{Security and privacy}

\keywords{property inference attack, differential privacy, federated learning, speech emotion recognition}



\renewcommand\footnotetextcopyrightpermission[1]{}
\settopmatter{printacmref=false}
\maketitle

\section{Introduction}
Speech, the natural and comfortable mean in human communication~\cite{5514597}, contains significant hidden paralinguistic information, such as emotional cues~\cite{schuller2013computational}. Recognizing emotional states from speech is essential for machines to understand humans' speech and effectively interact with humans in many applications, such as speech assistants, mental health monitoring, education, etc~\cite{ingale2012speech,paper40,paper43}. In the realm of human-computer interaction, Speech Emotion Recognition (SER) has garnered significant attention, particularly due to the utilization of Deep Neural Networks (DNNs) featuring intricate model structures.~\cite{khalil2019speech}.

DNNs have been successfully used in multiple areas, like image processing and audio processing tasks, as their deep structures can capture effective abstract information for those prediction tasks. Training DNNs often require large-scale datasets, thereby numerous models parameters can be sufficiently learned. However, it is challenging to gather personal data due to strict data protection laws and privacy concerns \cite{gdpr}, resulting in a bottleneck of training DNNs on personal speech data.

Federated Learning (FL), proposed by Google, enables distributed model training to tackle this issue \cite{paper2}.
Within the realm of FL, the exchange of information between local devices and the central server during each training iteration is limited to transmitting model parameters or gradients rather than the actual data. Subsequently, the central server aggregates these local models to construct a unified global model that serves as the foundation for the subsequent training round. This approach helps to protect personal data while still enabling the model to learn from diverse data sources. Specifically, in applying FL for SER, clients gather users' speech data from individual mobile devices, perform local model training, and subsequently transmit the models to a central server.
However, recent research has revealed that FL may be susceptible to property inference attacks~\cite{zhao2023privacy}. Despite only transmitting the model parameters or gradients during FL, sensitive information can still be inferred, as demonstrated in several studies~\cite{paper24,paper27,paper25,dlg,paper28}. In this context, FL may not provide rigorous privacy protection.

Property inference refers to the attacker learning information about the subject's data property, which is unrelated to the original learning task~\cite{hu2022membership}. Notably, the property inference in this work aims to learn a subject's attribute rather than the global data property~\cite{hu2022membership}.
While it has been studied in natural language processing and computer vision \cite{paper27,paper25,dlg,paper28}, its impact on speech signals has not been well-investigated. The principle of property inference through survey has been used in various attack methods \cite{paper31} that are already known.  If the attackers use property inference to attack the system, the profiles of the users (such as age, race, and gender) are very easily stolen. To foster awareness within the speech community, the imperative lies in developing robust protective measures against such attacks. Regrettably, a scarcity of studies has been dedicated to comprehensively evaluating and comparing various safeguarding techniques.

This study aims to assess and compare the efficacy of diverse privacy protection mechanisms in the context of FL-based SER. Our contributions are as follows. First, we introduce a novel privacy preservation technique called Property-Indistinguishability (Pro-Ind), and present experimental findings obtained by implementing various privacy protection strategies to thwart potential attacks. Second, we provide an empirical analysis of the outcomes stemming from using existing protection methodologies \cite{paper37,paper35} in conjunction with our proposed Pro-Ind method within the realm of FL-based SER (FL-SER). Third, this work verifies the proposed approach on inference attacks of multiple types of properties, offering a comprehensive comparison and guidance for future research works. In sum, the novelty of our work is that we have proposed a scheme to protect property privacy in voice, and its effectiveness has been verified through experiments.

The rest of the paper is organized as follows: Section \,2 introduces the relevant research studies. Section\,3 describes the proposed method in this work. Section \,4 shows the experimental evaluation. Finally, this work is concluded, and potential future work is discussed in Section\,5.

\section{Related Work}
\subsection{Federated Learning for SER}
More recently, FL has been applied to SER for protecting users' privacy in several studies. The study in~\cite{latif2020federated} demonstrated good performance of FL for SER compared to the state-of-the-art, and FL was shown to enhance the performance of audiovisual emotion recognition in~\cite{simic2024enhancing}. An FL-based system was proposed to monitor users' mental health by analyzing their emotional states from facial expressions and speech signals in~\cite{chhikara2020federated}. 

\subsection{Privacy Protection in Federated Learning}
Despite the capability of FL to protect data by only uploading model parameters or gradients to the server, it has been shown that FL may still expose user information to malicious attacks, such as property inference attacks discussed in this work. Attribute inference attacks were proposed to successfully attack an FL system for SER in~\cite{feng2021attribute}. The authors in~\cite{paper36} further evaluated User-level differential privacy (UDP) on attribute inference attacks in a framework of FL for SER.
In~\cite{zhao2023privacy}, a feature attention mechanism was proposed to guide the SER models to focus on learning emotion-related representations, thereby hiding emotion-irrelevant attributes. Paillier homomorphic encryption was applied to FL in the task of SER. Therefore, data remains private during training~\cite{mohammadi2023secure}
A method to protect gender information from inference attacks was proposed in~\cite{tan2023general}.

\section{Preliminary Knowledge}
\subsection{Federated Learning}

As aforementioned, in the FL-based SER (FL-SER), clients employ their speech data to train localized models on individual devices and transmit them to the central server. In this setting, the server has a potential malicious adversary that is curious (adhering to the FL protocol but wanting to get some information from clients' gradients). The server is expected to perform inference attacks using snapshots of each user's model parameters. This assumption is also confirmed in the recent work \cite{paper30}.

\subsection{Property Attacks}
 
\begin{figure}[t]
    \centering
	\includegraphics[width=0.35\textwidth]{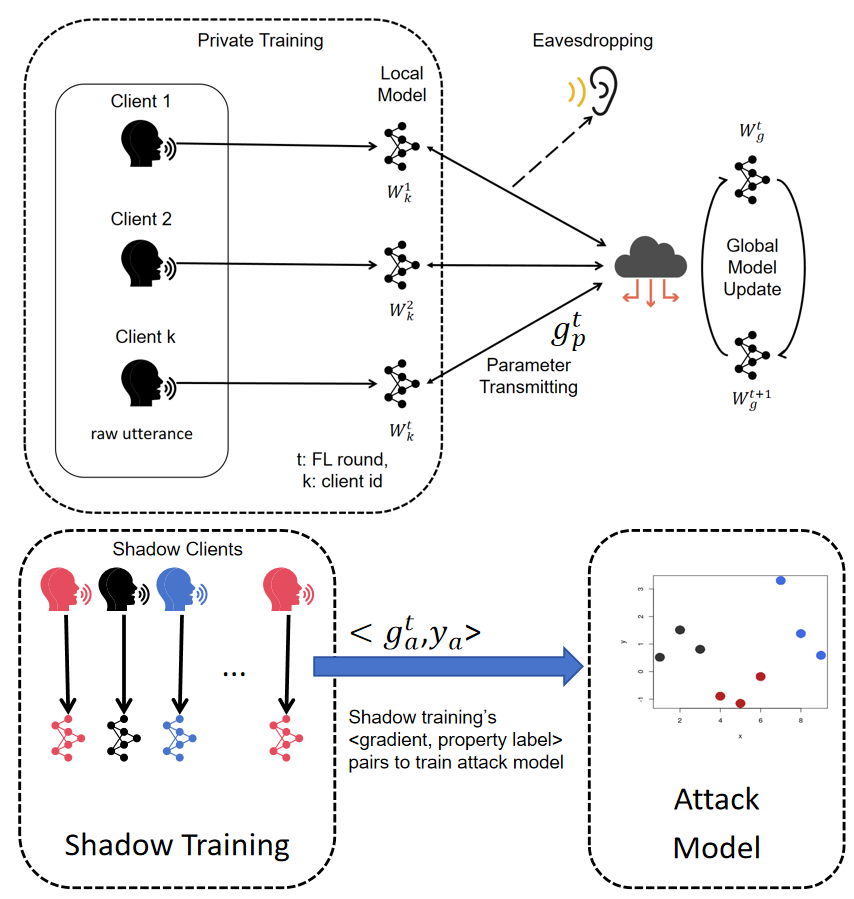}
 \vspace{10pt}
	\caption{The framework of attack.}
	\label{fig:before}
\end{figure}

We adopted the attack model shown in Figure \ref{fig:before}. This choice was made due to the model's standardization and ease of implementation. 
It's worth noting that attacks in the context of FL can be categorized into two main categories: those originating from within the system and those from external sources \cite{paper24}. Consequently, in the specific scenario of this paper, both the server, despite its honest but curious nature, and an external eavesdropper possess the capability to execute the attack. The above-mentioned framework outlining the attack process is visually represented in Figure \ref{fig:before}. This attack process has three distinct stages:

Firstly, \textbf{Private Training} is a procedure where the model's training takes place locally on the clients' devices. These clients engage with their localized training dataset, denoted as $\boldsymbol{D}_{\boldsymbol{p}}$, and associated emotional labels, $\boldsymbol{y}_{\boldsymbol{p}}$. 

The central server can coordinate the FL process through two primary algorithms: FedSGD and FedAvg \cite{paper44}. Only gradients or model parameters, rather than the raw data, are transmitted to the central server. In this setup, the gradients exposed by each client are represented as $\boldsymbol{g}^{\boldsymbol{t}}_{\boldsymbol{p}}$, where $\boldsymbol{t}$ specifies the training epoch and $\boldsymbol{p}$ identifies the client.

The central server coordinates the FL process using two main algorithms: FedSGD and FedAvg \cite{paper44}. Instead of transmitting raw data, only gradients or model parameters are sent to the central server. In this configuration, the gradients contributed by each client are denoted as $\boldsymbol{g}^{\boldsymbol{t}}_{\boldsymbol{p}}$, with $\boldsymbol{t}$ indicating the training epoch and $\boldsymbol{p}$ representing the client identifier.

Secondly, the concept of \noindent\textbf{Shadow Training} entails an attacker mimicking the client-side training process to obtain the training data necessary for an attack model. The attacker compiles datasets that are similar to the client's private dataset, $\boldsymbol{D}_{\boldsymbol{p}}$, to emulate the client-side training process. Consequently, gradients $\boldsymbol{g}^{\boldsymbol{t}}_{\boldsymbol{a}}$ are computed for each shadow client, denoted as $\boldsymbol{a}$, during training epoch $\boldsymbol{t}$, along with the respective property labels $\boldsymbol{y}_{\boldsymbol{a}}$.

Thirdly, \textbf{Attack Model} is denoted as $\boldsymbol{M}_{\boldsymbol{a}}$, the training dataset is comprised of gradients $\boldsymbol{g}^{\boldsymbol{t}}_{\boldsymbol{a}}$ and corresponding property labels $\boldsymbol{y}_{\boldsymbol{a}}$, both of which are acquired during the shadow training phase. The model is subsequently trained using these datasets and optimized to the parameter set $\Psi$. The objective is represented as $\mathcal{L}$:
%, following equation (\ref{for:attackModel}):

\begin{equation}
    \mathcal{L}=\min_{\Psi} \mathcal{L}(\boldsymbol{M}_{\boldsymbol{a}} (\boldsymbol{g}^{\boldsymbol{t}}_{\boldsymbol{a}},\Psi),\boldsymbol{y}_a).
    \label{for:attackModel}
\end{equation}

\noindent Upon the completion of the training process for the \textbf{Attack Model} $\boldsymbol{M}_{\boldsymbol{a}}$, it becomes feasible to predict the gender of client $\boldsymbol{p}$ by feeding the gradient $\boldsymbol{g}^{\boldsymbol{t}}_{\boldsymbol{p}}$ into the trained attack model.

\begin{figure}[t]
    \centering
	\includegraphics[width=0.35\textwidth]{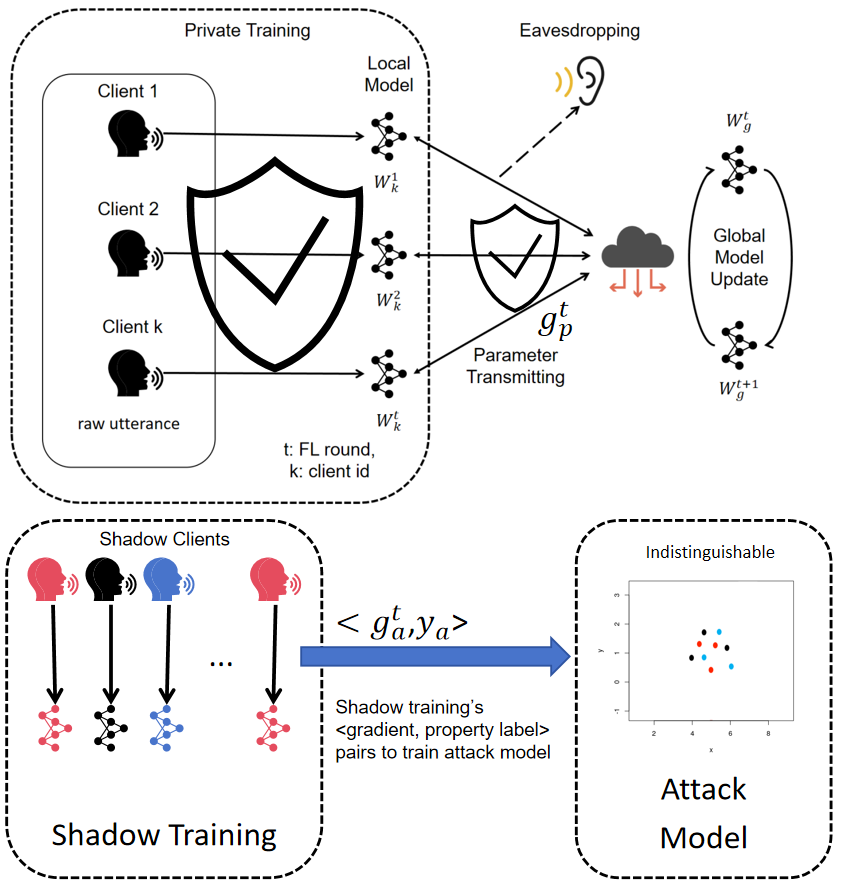}
 \vspace{10pt}
	\caption{The framework of protection.}
	\label{fig:after}
\end{figure}

\subsection{Protection Methods}

\noindent The principle of \textbf{Differential Privacy} ensures that the inclusion or omission of any single data point in a dataset does not substantially affect the conclusions drawn from the dataset. Formally, Differential Privacy is defined as per equation (\ref{for:classicalDP}) \cite{DP}. 
An randomized algorithm denoted as $\boldsymbol{M}$, which operates on a domain $\boldsymbol{N}^{\boldsymbol{|X|}}$ with a corresponding range $\boldsymbol{R}$, gives $(\boldsymbol{\epsilon}, \boldsymbol{\delta})$-differential privacy. This property holds true for any subset $\boldsymbol{S}$ within the range of $\boldsymbol{M}$ and for all adjacent data points $\boldsymbol{x}, \boldsymbol{x\prime} \in \boldsymbol{N}^{|\boldsymbol{X}|}$ as described below:

\begin{equation}
    \operatorname{Pr}[\boldsymbol{M}(\boldsymbol{x}) \in \boldsymbol{S}] \leq e^{\boldsymbol{\epsilon}} \operatorname{Pr}[\boldsymbol{M}(\boldsymbol{x\prime}) \in \boldsymbol{S}]+\boldsymbol{\delta}.
    \label{for:classicalDP}
\end{equation}

\noindent A comprehensive survey focusing on Differential Privacy (DP) for unstructured data has been presented in \cite{paper39}. This study uncovers that voiceprint\footnote{In this paper, we use X-vector-based speaker embedding to represent voiceprint.} combined with normal DP technique could be applied to speech data. However, as the principal objective of the current work is to safeguard client privacy against potential attacks—bearing in mind that the clients are actual human individuals—a more apt DP method, termed User-level DP, has been adopted for client-level privacy preservation. Figure \ref{fig:after} offers a graphical representation illustrating the modifications in the attack process when client-side protection measures are considered and incorporated.\\

\noindent\textbf{User-Level Differential Privacy (UDP):}
The primary concept of UDP is to introduce perturbation in local data using mechanism $\boldsymbol{M}_{\boldsymbol{u}}$. The perturbed data is ensured to be protected against privacy attacks with the help of parameter values $\boldsymbol{\epsilon}$ and $\boldsymbol{\delta}$. This protection technique is based on the research presented in Feng\emph{et al.}'s study \cite{paper36}.

\noindent In such a configuration, UDP introduces perturbations to the gradient $\boldsymbol{g}^{\boldsymbol{t}}_{\boldsymbol{k}}$ by appending Gaussian noise. This noise is characterized by a zero mean and a variance of $\boldsymbol{\sigma_{\boldsymbol{k}}^{2}} \boldsymbol{I}$ for a given gradient $\boldsymbol{g}^{\boldsymbol{t}}_{\boldsymbol{k}}$. The mathematical representation of this operation is detailed in equation (\ref{for:UDP}).

\begin{equation}
    \mathrm{\boldsymbol{M}_{\boldsymbol{u}}}(\boldsymbol{g}^{\boldsymbol{t}}_{\boldsymbol{k}})=\boldsymbol{g}^{\boldsymbol{t}}_{\boldsymbol{k}}+\boldsymbol{N}(0, \boldsymbol{\sigma_{\boldsymbol{k}}^{2}} \boldsymbol{I}).
    \label{for:UDP}
\end{equation}

\noindent The algorithmic specifics for UDP are elaborated upon in existing research \cite{paper36}. For a particular client, denoted as $\boldsymbol{k}$, the variance parameter $\boldsymbol{\sigma}_{\boldsymbol{k}}$ is delineated using equation (\ref{for:udpSigma}). The $\nabla{l}$ denotes the constraint on the gradient variance between consecutive datasets from a specific client $\boldsymbol{k}$ \cite{paper36}. Furthermore, $q$ and $T$ denote the proportion of client data sampled in each training epoch and the number of global training epochs, respectively. $\boldsymbol{\delta}_{\boldsymbol{k}}$ and $\boldsymbol{\epsilon}_{\boldsymbol{k}}$ are used to control differential privacy.

\begin{equation}
    \quad \quad  \boldsymbol{\sigma}_{\boldsymbol{k}} = \frac{\nabla l \sqrt{2 q T \ln \left(1 / \boldsymbol{\delta}_{\boldsymbol{k}}\right)}}{\boldsymbol{\epsilon}_{\boldsymbol{k}}}.
    \label{for:udpSigma}
\end{equation}

\noindent\textbf{voiceprint with DP}: operates on the foundational concept of differential privacy, yet its privacy definition is customized for voiceprints, specifically x-vectors \cite{snyder2018x}. To uphold privacy, noise is introduced into the voiceprint.

The voiceprint with DP approach, as described in \cite{paper37,ccs}, is specifically designed for the context of speech data publication. Given an initial data point $\boldsymbol{s}_{\boldsymbol{0}}$, the mechanism, denoted as $\boldsymbol{M}_{\boldsymbol{v}}$, initially extracts an x-vector $\boldsymbol{v}_{\boldsymbol{0}}$. Subsequently, this x-vector is transformed into a perturbed version, $\widetilde{\boldsymbol{v}}$, which belongs to the dataset $\boldsymbol{D}_{\boldsymbol{v}}$. The dataset $\boldsymbol{D}_{\boldsymbol{v}}$ consists of x-vectors extracted from a yet-to-be-published speech dataset $\boldsymbol{D}_{\boldsymbol{s}}$. The transition from $\boldsymbol{v}_{\boldsymbol{0}}$ to $\widetilde{\boldsymbol{v}}$ occurs with a probability that is a function of the distance between these two vectors. Finally, the perturbed x-vector $\widetilde{\boldsymbol{v}}$ is combined with the original speech data $\boldsymbol{s}_{\boldsymbol{0}}$ to synthesize a modified speech data point, $\widetilde{\boldsymbol{s}}$, which is then released for publication.

\section{The Proposed Approach: Property-Indistinguishability}
While voiceprints contain property information, every property information is coupled together \cite{Raj2019ProbingTI}. Individual users exhibit varying privacy requirements in the FL context. For instance, while some users may wish to withhold information about their age, others may prefer to conceal data related to their gender or ethnicity. Apart from these specific information that clients choose not to disclose, they are generally amenable to contributing other personal information for FL tasks. Consequently, utilizing voiceprint with DP as a mechanism to thwart property inference attacks is suboptimal. This is because the very act of preserving a particular facet of privacy necessitates obfuscating that specific data. When clients engage in specific FL applications, such as SER, they must furnish information about vocal characteristics and emotional states. The intentional obfuscation of this data for privacy protection could potentially impede the optimization of FL accuracy. As a result, we propose a new protection method called Pro-Ind, which is specifically designed for property privacy preservation.

\begin{definition}[Property-Indistinguishability, i.e., Pro-Ind]
\noindent A mechanism $\boldsymbol{M}_{\boldsymbol{g}}$ is said to fulfill $\boldsymbol{\epsilon}$-property-indistinguishability under the following conditions: For any resulting property embedding $\widetilde{\boldsymbol{c}}$ and any two feasible input embeddings $\boldsymbol{c}, \boldsymbol{c}\prime \in \boldsymbol{C}$ as follows:

\begin{equation}
    \frac{\operatorname{Pr}(\mathcal{M}_g(c) = \Tilde{c})}{\operatorname{Pr}(\mathcal{M}_g(c') = \Tilde{c})} \leq e^{\boldsymbol{\epsilon} \boldsymbol{d}_{\mathcal{C}}(\boldsymbol{c}, \boldsymbol{c} \prime)}.
\end{equation}

\noindent where $\boldsymbol{C}$ denotes the set of properties represented in the forms of embeddings. The function $\boldsymbol{d}_{\boldsymbol{C}}$ quantifies the distance between angulars of two embeddings $\boldsymbol{c}$ and $\boldsymbol{c}\prime$.
\end{definition}

It is reasonable to suppose that clients should possess equivalent capabilities, especially considering that an adversary could potentially leverage property embeddings from publicly accessible datasets.

\subsection{Property Embedding Protection under Pro-Ind}
\noindent The method proposed in this paper, termed Pro-Ind, is architected specifically to safeguard property-related information. This is achieved by injecting noise into the property embeddings derived from an x-vector framework trained using property identifiers rather than speaker identifiers. Unlike the data release scenario elaborated in \cite{paper37,ccs}, the present framework is engineered for application within FL. It accommodates the client's use of public datasets, mirroring the adversary's capabilities and thus reinforcing privacy protection.

With an initial property embedding denoted as $\boldsymbol{c}_{\mathbf{0}}$, the mechanism $\boldsymbol{M}_{\boldsymbol{g}}$ introduces perturbations by probabilistically selecting a property embedding $\widetilde{\boldsymbol{c}}$ from the dataset $\boldsymbol{D}_{\boldsymbol{g}}$. The selection process is governed by predefined probability distributions, thereby affording a degree of plausible deniability for the original property embedding $\boldsymbol{c}_{\mathbf{0}}$. Their definitions are as follows:

\begin{equation}
    \operatorname{Pr}(\mathcal{M}_g(c_0)= \Tilde{c}) \propto e^{-\boldsymbol{\epsilon} \boldsymbol{d}_{\mathcal{C}}\left(\widetilde{\boldsymbol{c}}, \boldsymbol{c}_{0}\right)}.
\end{equation}

\begin{algorithm}\small
\SetKwInput{KwInput}{Input}                % Set the Input
\SetKwInput{KwOutput}{Output}              % set the Output
\caption{Protecting Property-Embedding}
\DontPrintSemicolon
  
\KwInput{Speech sentence $s_0$, Dataset $D_g$, Property embedding extracting model $M_E$, Speech Synthesize model $M_S$}
\KwOutput{Protected speech $\tilde{s}$}
  %\KwData{Testing set $x$}
$c_0$ $\longleftarrow$ $M_E$($s_0$)\;

\For{each $s_i$ from $D_g$:}{
    $c_i$ $\longleftarrow$ $M_E$($s_i$)\;
    
    Compute $\operatorname{Pr}(\mathcal{M}_g(c_0)= c_i)$
}

Select property embedding $\tilde{c}$ randomly according to $\operatorname{Pr}(\mathcal{M}_g(c_0)= c_i)$

$\tilde{s}$ $\longleftarrow$ $M_S$($\tilde{c}$, $s_0$)\;

\Return{$\tilde{s}$}
\end{algorithm}

\noindent The synthesis model employed in this work is an end-to-end neural network, trained by utilizing the ESPnet (End-to-end Speech Processing Toolkit) framework \cite{hayashi2020espnet}, specifically the example setup in egs/libritts/tts1. This model is designed to generate Mel Spectrogram (Mel-spec) features \cite{mel-spec}, which serve as the input for a waveform vocoder \cite{vocoder}.

\subsection{Privacy Analysis}

\noindent The crux of privacy leakage resides in the gap between an adversary's prior and posterior knowledge concerning a targeted individual. Although Differential Privacy (DP) is not capable of entirely thwarting adversaries from accumulating knowledge, it can impede their progress through the judicious selection of parameters (the detailed description can be found in \cite{DP}).

Given two distinct property embeddings, $\boldsymbol{c}$ and $\boldsymbol{c}'$, and an output property embedding, $\tilde{\boldsymbol{c}}$, generated by the protective mechanism:

\begin{align}
    \frac{\operatorname{Pr}(\mathcal{M}_g(c)= \Tilde{c})}{\operatorname{Pr}(\mathcal{M}_g(c')= \Tilde{c})} &= \frac{e^{-\boldsymbol{\epsilon} \boldsymbol{d}_{\mathcal{C}}\left(\widetilde{\boldsymbol{c}}, \boldsymbol{c}\right)}}{e^{-\boldsymbol{\epsilon} \boldsymbol{d}_{\mathcal{C}}\left(\widetilde{\boldsymbol{c}}, \boldsymbol{c}'\right)}} \nonumber \\ 
    &= e^{\boldsymbol{\epsilon} \boldsymbol{d}_{\mathcal{C}}\left(\widetilde{\boldsymbol{c}}, \boldsymbol{c'}\right) - \boldsymbol{d}_{\mathcal{C}}\left(\widetilde{\boldsymbol{c}}, \boldsymbol{c}\right)},
\end{align}
    
\noindent because the distance metric $\boldsymbol{d}_{\mathcal{C}}(\boldsymbol{c}, \boldsymbol{c'})$ is the angular distance, which satisfies the triangle inequality, so there are definitions as follows:
\begin{equation}
    \boldsymbol{d}_{\mathcal{C}}(\boldsymbol{\Tilde{c}}, \boldsymbol{c'}) - \boldsymbol{d}_{\mathcal{C}}(\boldsymbol{\Tilde{c}}, \boldsymbol{c}) \leq \boldsymbol{d}_{\mathcal{C}}(\boldsymbol{c}, \boldsymbol{c'}),
\end{equation}
\noindent therefore

\begin{equation}
    \frac{\operatorname{Pr}(\mathcal{M}_g(c)= \Tilde{c})}{\operatorname{Pr}(\mathcal{M}_g(c')= \Tilde{c})} \leq e^{\epsilon \boldsymbol{d}_{\mathcal{C}}(\boldsymbol{c}, \boldsymbol{c'})},
\end{equation}
\noindent which implies that the property embedding protection method complies with the Pro-Ind criterion.

\section{Experiments}
\subsection{Data Description}

\noindent The CREMA-D database \cite{database} contains 7,442 original audio clips provided by 91 actors. The actor demographics are varied, with 48 male and 43 female participants representing diverse racial and ethnic backgrounds, including African American, Asian, Caucasian, Hispanic, and Unspecified. The age range of the contributors spans from 20 to 74 years. Each clip includes one of twelve distinct sentences and is annotated with mood level (in total four levels) and emotional state (in total six states: sad, neutral, happy, fear, disgust, angry). For the present experiment, we focused on four specific elemental emotional categories: neutral, pleased, joyful, and furious.

\textbf{Evaluation Metrics.} In this paper, we use multiple metrics for evaluating the results, including Accuracy ($ACC$), Unweighted Average Recall ($UAR$), Success Rate of Property Attack ($SR_P$), and Unweighted Average Success Recall of Property Attack ($UASR_P$). Their calculation procedures are in the following.

\begin{equation}
    ACC = \sum_i^{i \in E} \frac{\hat N^E_i}{\tilde N^E_i}  r^E_i 
\end{equation}

\begin{equation}
UAR = \frac{1}{N^E} \sum_i^{i \in E} \frac{\hat N^E_i}{N^E_i},
\end{equation}

\begin{equation}
SR_P = \sum_j^{j \in P} {\frac{\hat N^P_j}{\tilde N^P_j}  r^P_j},
\end{equation}

\begin{equation}
UASR_P = \frac{1}{{N^P}}\sum_j^{j \in P} \frac{\hat N^P_j}{N^P_j},
\end{equation}
where $E$ denotes the $N^E$ emotional classes, $N^E_i$ is the number of samples in the $i$-th emotional class, $\hat N^E_i$ indicates the number of correctly classified samples, $\tilde N^E_i$ is number of predicted class at the $i$-th index, and $r^E_i$ is the ratio of the $i$-th class in the whole dataset. Correspondingly, $P$ means the $N^P$ property groups, $N^P_j$ is number of samples in the $j$-th property class, $\hat N^P_j$ means number of samples with correctly predicted properties, $\tilde N^P_j$ is number of predicted property class at the $j$-th index, and $r^P_j$ is the ratio of the $j$-th property class.

\begin{figure*}[]
\centering

\subfigure[Age Attack with Each Protection]{
\includegraphics[scale=0.17]{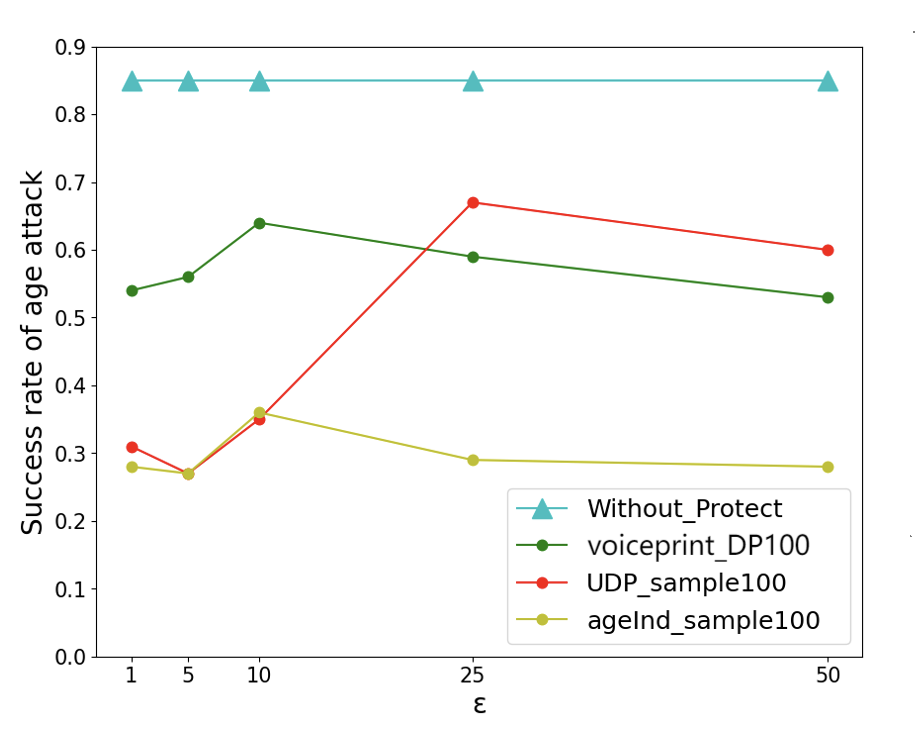}
}%
%\\
\subfigure[Race Attack with Each Protection]{
\includegraphics[scale=0.17]{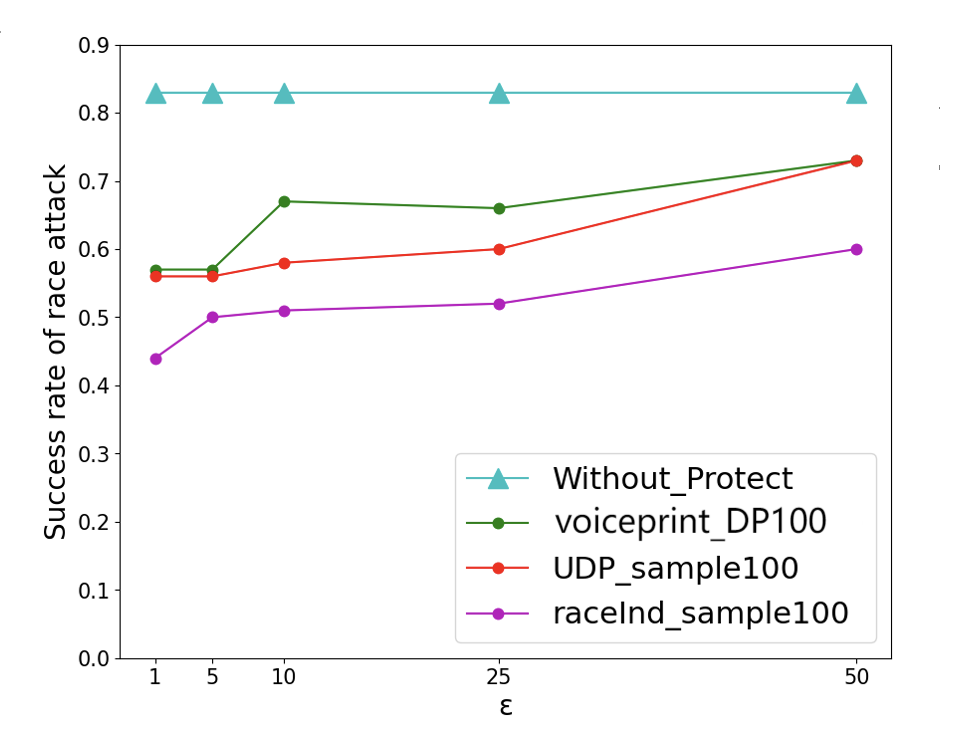}
}%
%\\

\subfigure[Gender Attack with Each Protection]{
\includegraphics[scale=0.17]{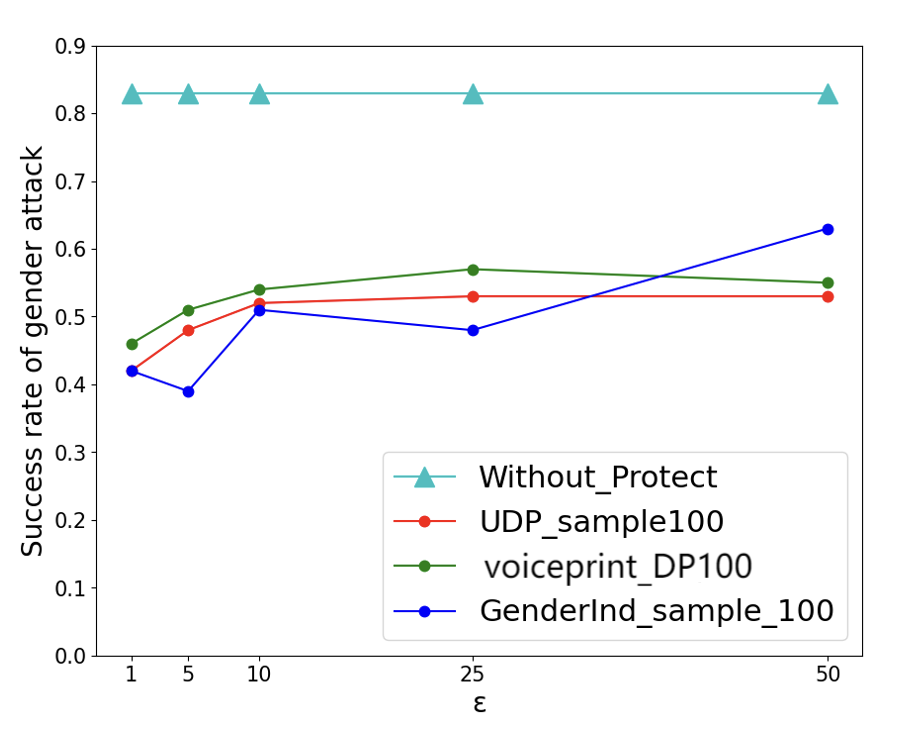}
}%
\caption{The success rates of attack models (age, gender, race) with corresponding Pro-Ind}
\label{fig:proIndAtt}
\end{figure*}

\begin{figure}[]
    \centering
	\includegraphics[width=0.36\textwidth]{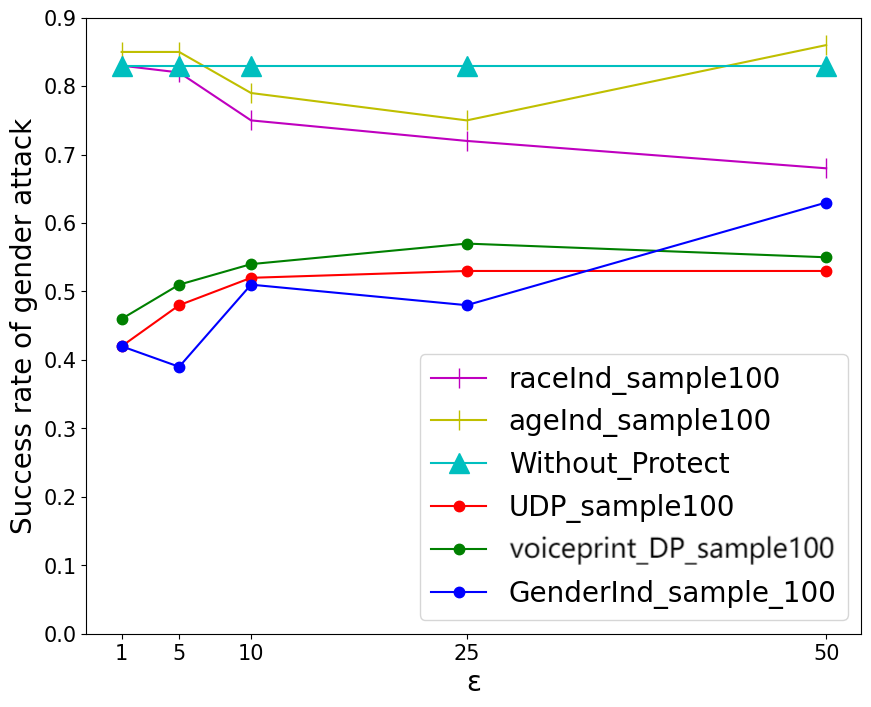}
	\caption{The success rate of the gender attack model with different Pro-Ind}
	\label{fig:genderIndAtt}
\end{figure}

\begin{figure}[]
    \centering
	\includegraphics[width=0.4\textwidth]{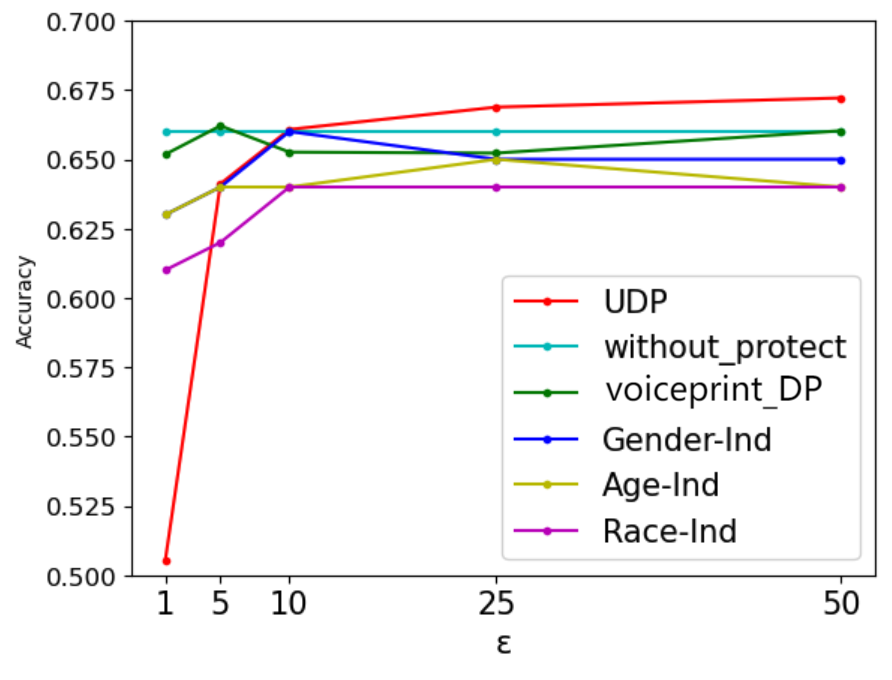}
 \vspace{10pt}
	\caption{The accuracy of FL-SER model.}
	\label{fig:accFL}
\end{figure}

\subsection{Results of Unprotected FL}

\noindent The private training methodology deployed in this study is consistent with approaches presented in prior research \cite{paper36}. The DNN architecture designated for SER is structured with two dense layers, with hidden layer dimensions of 256 and 128 units, respectively. The activation function incorporated within the network is the Rectified Linear Unit (ReLU), and a dropout rate of 0.2 is implemented to mitigate overfitting. For the FL coordination, the most popular FedAvg algorithm is used as the foundational framework.

\begin{table*}[ht]%\small

\begin{center}

\caption{Experimental results (The attack success rates and FL-SER model accuracy without protections). Fold means subsets of training data with the random factor.)}

\label{tab:accattack}
\begin{tabular}{|c|c|c|c|c|c|c|c|c|}
\hline
 & \multicolumn{2}{c|}{Gender Attack model} & \multicolumn{2}{c|}{Age Attack model} &
 \multicolumn{2}{c|}{Race Attack model} & \multicolumn{2}{c|}{FL-SER model} \\
 \cline{2-9}
 & $SR_P$ & $UASR_P$ & $SR_P$ & $UASR_P$ & $UASR_P$& $UASR_P$ & $ACC$ & $UAR$\\
\hline
Fold1 & $0.84$ & $0.83$ & $0.85$ & $0.84$ & $0.82$ & $0.85$ & $0.66$ & $0.60$ \\
\hline
Fold2 & $0.85$ & $0.84$ & $0.85$ & $0.86$ & $0.83$ & $0.80$ & $0.67$ & $0.60$ \\
\hline
Fold3 & $0.82$ & $0.80$ & $0.86$ & $0.87$ & $0.85$ & $0.78$ & $0.66$ & $0.62$ \\
\hline
\end{tabular}
%\end{center}

\end{center}
\end{table*}

\noindent In the shadow training phase, a total of 30 speakers are employed, with the remaining speakers allocated for private training. The client participation rate in private training is set at 10\%. For each client, 80\% of the available data is designated for local training, and the remaining 20\% is reserved for validation purposes. The local learning rate is established at 0.0005, the number of local epochs is fixed at 1, and the total number of global training epochs, denoted as \( T \), is set at 200. Furthermore, the local training batch size is configured to be 20.

Aligned with the configurations in prior work \cite{paper36}, the architecture remains constant, and additional layer experimentation is foregone. This choice is guided by the understanding that the first layer is the most susceptible to privacy leakage \cite{paper30}. We also adhere to the default configuration for voiceprint with DP as in \cite{paper37,ccs}.

To assess the efficacy of various defense strategies, multiple values of differential privacy's \( \boldsymbol{\epsilon} \) are tested, specifically \( \{1, 5, 10, 25, 50\} \). 100 gradients are extracted to assess the performance of the attack model. Table \ref{tab:accattack} presents detailed performance for both the FL-SER model and the attack model without a protection mechanism.

\subsection{Results of Protected FL}
\noindent In parallel to the methodologies proposed in Feng's research \cite{paper31}, this work expands the scope of investigation by designing two additional types of attack models, specifically focusing on age and race. For each variety of attack, three distinct defense mechanisms—UDP, voiceprint with DP, and Pro-Ind—are employed, each tailored to defend against the attributes targeted by the attack model in question. As illustrated in Figure \ref{fig:proIndAtt}, Pro-Ind ostensibly outperforms the other defense strategies in safeguarding privacy. It is noteworthy to mention that in the absence of any defense mechanisms, the success rate for each type of attack model hovers around \(83\% - 85\%\), as summarized in Table \ref{tab:accattack}.

\noindent Figure \ref{fig:accFL} reveals a nuanced impact of employing various privacy-preserving techniques on the accuracy of the FL model. Notably, the use of voiceprint with DP and Pro-Ind incurs a marginal degradation in model performance, affirming their efficiency as privacy-preserving methods with minimal trade-offs in accuracy. Conversely, the employment of UDP demonstrates a more pronounced impact on the FL model, especially when the parameter \(\boldsymbol{\epsilon}\) is set to a smaller value, implying stronger privacy protections. Under such circumstances, the model's performance experiences a significant decline. For a comprehensive understanding, the baseline accuracy of the FL model without any privacy-preserving techniques is cataloged in Table \ref{tab:accattack}.

On the other hand, the experiment also compares the privacy protection of UDP, voiceprint with DP, pro-Ind for gender, pro-Ind for age, and pro-Ind for race against gender attacks. As shown in Figure \ref{fig:genderIndAtt}, it can be seen that good results will not be achieved when misusing non-corresponding pro-Ind. This interesting result also indirectly proves the accuracy of pro-Ind from the side.

\noindent Figures \ref{fig:proIndAtt} and \ref{fig:genderIndAtt} present comparative evaluations of the efficacy of various privacy-preserving techniques, namely, UDP, voiceprint with DP, and Pro-Ind. Among these, Pro-Ind emerges as the most proficient method for safeguarding user privacy without significantly compromising the performance of the FL model.

Specifically, UDP introduces noise at the gradient level, thereby obfuscating information globally across all users. Voiceprint with DP, on the other hand, perturbs only the voiceprint data. In contrast, Pro-Ind focuses on adding noise to attribute embeddings. This targeted approach allows Pro-Ind to offer a more nuanced balance, ensuring robust privacy protection while minimally affecting the utility of the FL model.

\subsection{Discussion}

\noindent It is critical to design and deploy privacy-preserving machine learning algorithms in distributed computing environments like FL.
In light of the experimental findings elucidated, we can conclusively assert that to preserve specific facets of privacy within the domain of speech-emotion FL, the judicious strategy involves explicitly defining the specific attributes to be protected and safeguarding property-relevant embeddings through mechanisms like Property-Indistinguishability. The proposed approach can effectively prevent the attackers from getting the user profile. Deploying such targeted privacy-preservation mechanisms demonstrates advantages over a blanket approach of indiscriminate noise addition to the entire model. This specialized method fosters enhanced accuracy in the FL model and ensures more effective privacy preservation. Consequently, it represents an optimal trade-off between utility and privacy.

\section{Conclusion and Future Work}
\label{sec:conclude}
\noindent This paper proposed a novel privacy-preserving method within the context of Federated Learning-based Speech Emotion Recognition (FL-SER). Empirical evidence substantiated that more effective privacy safeguards and improved FL-SER accuracy are realized when privacy protection mechanisms are architected in alignment with precisely tailored privacy definitions. This insight provided a fertile ground for subsequent research endeavors to develop nuanced definitions for speech privacy. In extending this line of inquiry, we postulate a broader hypothesis: that the alignment of protection mechanisms with well-articulated privacy definitions is imperative for achieving the pinnacle of privacy protection and FL model accuracy, not just in the realm of FL-SER but across the entire spectrum of speech-related FL tasks. Future research will undertake further empirical investigations into a diverse array of speech-related tasks to corroborate the validity of this hypothesis, especially in protecting user privacy in voice data in Large Language Models like ChatGPT.

%\begin{acks}
%To Robert, for the bagels and explaining CMYK and color spaces.
%\end{acks}

%\newpage
\bibliographystyle{ACM-Reference-Format}
\bibliography{paper}

\iffalse
\appendix

\section{Research Methods}

\subsection{Part One}

Lorem ipsum dolor sit amet, consectetur adipiscing elit. Morbi
malesuada, quam in pulvinar varius, metus nunc fermentum urna, id
sollicitudin purus odio sit amet enim. Aliquam ullamcorper eu ipsum
vel mollis. Curabitur quis dictum nisl. Phasellus vel semper risus, et
lacinia dolor. Integer ultricies commodo sem nec semper.

\subsection{Part Two}

Etiam commodo feugiat nisl pulvinar pellentesque. Etiam auctor sodales
ligula, non varius nibh pulvinar semper. Suspendisse nec lectus non
ipsum convallis congue hendrerit vitae sapien. Donec at laoreet
eros. Vivamus non purus placerat, scelerisque diam eu, cursus
ante. Etiam aliquam tortor auctor efficitur mattis.

\section{Online Resources}

Nam id fermentum dui. Suspendisse sagittis tortor a nulla mollis, in
pulvinar ex pretium. Sed interdum orci quis metus euismod, et sagittis
enim maximus. Vestibulum gravida massa ut felis suscipit
congue. Quisque mattis elit a risus ultrices commodo venenatis eget
dui. Etiam sagittis eleifend elementum.

Nam interdum magna at lectus dignissim, ac dignissim lorem
rhoncus. Maecenas eu arcu ac neque placerat aliquam. Nunc pulvinar
massa et mattis lacinia.
\fi

\end{document}